# The Missing Link: Identifying Digital Intermediaries in E-Government[1]

Sergio Toro-Maureira, Universidad Mayor, Chile

Alejandro Olivares, Universidad Mayor, Chile

Rocío Sáez-Vergara, Millennium Nucleus on Digital Inequalities and Opportunities, Chile

Sebastián Valenzuela, Pontificia Universidad Católica de Chile, Chile

Macarena Valenzuela, Universidad Católica de Temuco, Chile

Teresa Correa, Universidad Diego Portales, Chile

November 19, 2024

**UNDER REVIEW**

## Abstract

The digitalization of public administration has advanced significantly on a global scale. Many governments now view digital platforms as essential for improving the delivery of public services and fostering direct communication between citizens and public institutions. However, this view overlooks the role played by 'digital intermediaries'—agents who, while not formally part of the government, significantly shape the provision of e-government services. Using Chile as a case study, we analyze these intermediaries through a national survey on digitalization. We find five types of intermediaries: family members, peers, political figures, bureaucrats, and community leaders. The first two classes comprise 'close' intermediaries, while the latter three comprise 'hierarchical' intermediaries. Our findings suggest that all these intermediaries are a critical but underexplored element in the digitalization of public administration.

Digitalization has become somewhat of an inevitable path for public administrations worldwide. According to a World Bank study (2021), 80 out of 198 countries under study (GovTech) and it was observed that 43 economies present a maturity digital government (Nii-Aponsah et al., 2021). This transformation poses significant challenges for governments, as they must integrate increasingly sophisticated digital systems into their operations (Meijer and Zouridis, 2004; Ragulina et al., 2021; Torres et al., 2005; Wimmer and Traunmüller, 2000).

The literature indicates that government digitalization positively affects the administration of public benefits through greater efficiency and access. These effects are observed both in the design of processes and the streamlining of procedures, as well as in the transparency of allocations and the type of linkage with beneficiaries (Abdou, 2021; Androniceanu et al., 2022; Cordella, 2007; Cordella and Tempini, 2015; Hazlett and Hill, 2003; Henman, 2010; Mendoza and Delgado, 2022). For example, it is assumed that digitalizing public services favors direct and unmediated relationships between citizens and their governments. It is argued almost axiomatically that digitalization has allowed people to obtain benefits directly without the need for face-to-face interactions and that this transformation, consequently, generates a 'virtuous cycle' because it provides a direct link with less discretionary and bureaucratic burden, in addition to better limiting clientelism in the allocation of benefits (Bellamy and Taylor, 1998; Cordella, 2007; Gil-García and Pardo, 2005; Morozov, 2017; Osborne and Plastrik, 1997).

To examine these arguments, the purpose of this study is to propose a typology of intermediaries using two major dimensions which is tested using the 2023 Digital Inclusion Survey (Correa & Pavez 2023) surrounding the direct relationship between e-government and public service beneficiaries. We identify intermediary actors whose presence complicates the assumed linear connection between governments and citizens. These intermediaries, though not formally part of the service delivery process, play a crucial role for beneficiaries who lack the digital abilities or connectivity to access e-services independently. We refer to these actors as 'digital intermediaries.'

To examine these arguments, the purpose of this study is to propose a typology of intermediaries using two major dimensions (hierarchical and close) involving family actors and political agents. Furthermore, within each dimension lie sub-dimensions that include the proximity to beneficiaries and the procedural complexity. We used the 2023 Digital Inclusion Survey (Correa & Pavez, 2023) developed in Chile. With a 93% internet penetration rate and significant progress in government digitalization, Chile is a crucial case for constructing this typology.

This article is organized into four sections. The first section offers a theoretical overview, analyzing the literature on e-government dynamics and the benefits of public service digitalization. The second section explores the concept of digital intermediaries, examining the shifts and continuities in the government-citizen relationship in the context of digitalized public services. The third section develops a typology of digital intermediaries, applies factor analysis techniques to test questions related to digital intermediation, and proposes dimensions and sub-dimensions for future research. The article concludes with a discussion on the scope and implications of the debate surrounding digital intermediaries.

## Government Digitalization: Gaps and Intermediation

Research on government digitalization emerged in the early 2000s. These studies indicate that e-government mechanisms significantly improve public services by increasing accessibility, efficiency, and transparency (Beynon-Davies, 2005; Saha, 2009; Drigas, et al., 2005; Vatuiu, 2008; Von Haldenwang, 2004; Weerakkody, et al., 2006). Similarly, other research argues that digitalization improves bureaucratic procedures seeking citizen-centered services with direct interactions and no intermediaries (Ljungholm, 2015; Torres, et al., 2005; Monga, 2008; Vyas-Doorgapersad, 2009). Meanwhile, Brown (2005), Torres et al. (2005), and Cordella (2007) show that the new technologies positively impact public administration through greater efficiency in government operations.

However, direct interaction through digital technologies has added and complicated some tasks and objectives of traditional bureaucracy. Indeed, some scholars scrutinize the benefits of digitalization highlighted by the public administration literature. Buffat (2015), for example, analyzes the dual effects of technologies at the administrative level and in the deployment of public services (Also called street-level bureaucracy). He challenges the potential reduction of front-line bureaucratic discretion, termed the curtailment thesis, and the enhancement of capabilities through additional resources, known as the enablement thesis. The curtailment thesis suggests that e-government technologies might limit bureaucratic discretion by standardizing procedural processes and reducing individual judgment. In contrast, the enablement thesis postulates that these technologies might empower officials and citizens by facilitating access to information, improving communication and optimizing service delivery. Buffat (2015) notes that technology is reshaping the landscape of public administration and interactions between citizens and governments and cautions about the problems of digital deployment and the real effectiveness of this format in achieving public objectives.

In the same vein, Jansson and Erlingsson (2014) address the difficulties of transitioning from a traditional bureaucracy to a digital one, particularly the shift from traditional public administration, where street-level bureaucrats build legitimacy through direct interactions with citizens, to e-government, which seems to distance those relationships. Jansson and Erlingsson (2014) point out that e-government is a double-edged sword, as it promises greater efficiency but undermines the relational dynamics between bureaucrats and citizens. More simply stated, e-government streamlines procedures while reducing the government's understanding of and closeness to citizens.

While this literature does not seek a return to the traditional bureaucracy, it does warn of the challenges and problems of an e-government design that has developed without self-reflection in most countries (Lenk and Traunmuller, 2000). A review of theoretical and empirical analyses presents at least three issues governments must face when delivering services digitally: the connectivity gap, the digital literacy gap, and the change in intermediation to access public services.

The first issue is the connectivity gap. In fact, lack of access is one of the main difficulties reported by both scholars and practitioners. Global research shows that a significant part of the world's population is on the margins of digitalization. This connectivity gap hinders the intended distribution of public benefits because the population is not reached homogeneously (Rodríguez Fernandez and Gutierrez, 2017; Tully, 2003; Tully, 2007). Widespread reports reveal that about 2.6 billion people in the world remain disconnected, most of them in rural areas.[2] So government digitalization would be different for populations in areas without connectivity (Sovetova, 2011). Some scholars have suggested moving towards connectivity technologies better suited to rural environments. Tognisse and Degila (2021), for example, point out that the implementation of telecommunication services has been a predominantly urban activity because there are more apparent returns on investment. They suggest, therefore, an expansion of mobile networks and adopting technologies tailored to these needs. Similarly, Khalil et al (2019) test the suitability of specific technologies in rural areas to provide global internet access. Khalil et al's study is pioneering as it compares different network architectures and performance parameters, focusing on how these technologies can be effectively deployed in rural environments to mitigate the digital divide.

The second issue is the digital literacy gap. Digital literacy is the ability of individuals to understand technologies for complex purposes and to use devices to solve needs. It is argued that this relationship with technology is as important as connectivity (Calderón, 2019; Erstad, 2011; Schäfer, 2011). Numerous studies have identified factors that play critical roles in the development of digital literacy. Namely, they are age (Abad-Alcalá, 2014; Calderón, 2019; Campo Sánchez and Mancilla, 2015; Fernández and Gutiérrez, 2017; Bordelba and Garreta, 2018; Tully, 2003; Tully, 2007), education (Haight et al., 2016; Dutton and Reisdorf, 2019; Schäfer, 2011), and economic status (Calderón, 2019; Fernández and Gutiérrez, 2017; Tully, 2007). More specifically, lags in digital literacy, like other individual skill development processes, strongly correlate with older age, lower economic status (possibly due to lack of connectivity), and lower educational level (possibly due to lack of prerequisite skills like general literacy).

The third issue is the change in intermediation to access public services. This results from gaps in digital literacy and connectivity, as well as how governments connect with beneficiaries through e-government platforms. Indeed, replacing the physical service office process with a digital process wholly changes the dynamics and the identities of the intermediating actors. More specifically, simplifying bureaucratic procedures activates intermediaries external to the government. Although there is little research about digital intermediaries in the literature, the concept of digital intermediaries has been discussed since the beginning of government digitalization. Indeed, as early as 2004, Pasic argued that intermediaries are facilitating agents of access to public services, improving the efficiency of service delivery. Pasic et al (2004) highlight the role of intermediaries in integrating inter-agency services through mechanisms adapted to users' needs. Similarly, Griffin and Halpin (2002) look at the local government context to note that digital intermediaries create trusted environments that enable digitalization of public services with greater citizen engagement. In this regard, digital intermediaries are fundamental to the success of e-government and the provision of public services. However, technological advances and digital massification have made the identification of digital intermediaries more complex. In the following section, we propose a typology to help understand the phenomenon of intermediation.

## Digital Intermediaries: A Typology

Governments usually allocate public benefits to optimize their effectiveness. Before digitalization, allocation was solely based on the volume of forms received at respective service offices. The public administration literature uses traditional and street-level bureaucracy concepts to define these scenarios. Both types of bureaucracy are composed of public officials responsible for intermediating between the government and the citizenry, processing applications to enter social benefit systems in the field or in service offices (Hill, 2003; Lipsky, 1971, 1976, 2010; Maynard-Moody and Musheno, 2000; Maynard-Moody and Portillo, 2010; Prottas, 1978; Ricucci, 2005; Snellen, 2002).

The bureaucrats in charge have the discretion to adapt policy regulations to specific situations or manage allocation within the population (Hupe and Buffat, 2014; Hupe and Hill, 2003; Lipsky; 2010; Maynard-Moody and Musheno, 2000; Maynard-Moody and Portillo, 2010; Prottas, 1978; Tummers and Bekkers, 2014). These bureaucrats act as intermediaries between the government and the people, addressing implementation gaps caused by the complexity of benefits and the number of actors involved in the process (Buchely, 2015; Leyton, 2023). When there are obstacles to accessing government aid, bureaucratic intermediaries act by approaching, translating, interpreting, or carrying out the formalities for citizens. Their intermediation has been vital in implementing and applying policies to diverse social environments, using their professional expertise and discretion based on their knowledge of the requirements.

Both traditional bureaucracy and street-level bureaucracy operate in arcs of action regulated and known by all the actors in the process (Casas Arango, Aguirre Henao and Mancilla López, 2021; Leyton, 2023; Lipsky, 2010; Navarrete and Figueroa, 2019). The literature has studied the linkages between intermediaries and recipients, which vary in modality and are formed from socially dissimilar actors who interact without seeking reciprocity of information or benefits, creating heterogeneous intermediations (Hamilton, Hileman, and Bodin, 2020).

Government digitalization has transformed these dynamics, shifting intermediation purely by bureaucratic actors to others outside the official government process that support beneficiaries without being formally included in the design of services. Based on this finding, we develop a typology of intermediation that includes different agents that can intervene in public procedures. This digital intermediation will be classified as hierarchical and close, as this article shows in figure 1 with some examples.

Figure 1. Typology of digital intermediary-user relationships

| Intermediation ||
|---|---|
| **Hierarchical** | **Close** |

| Political | Bureaucratic | Community | Family | Peer-to-peer |
|---|---|---|---|---|
| -political agent (local and national) | -state official | -neighborhood leaders<br>-teacher | -parent<br>- son/daughter | -spouse/partner<br>-friend<br>-neighbor |

**Hierarchical Digital Intermediation**

Hierarchical digital intermediaries are vertical links with legal or symbolic authority outside the individual's immediate circle. In this category, we place politicians, public servants, and social leaders with better information-handling skills to manage online procedures. Within the category of hierarchical digital intermediaries, we identify three subtypes: political intermediaries, bureaucratic intermediaries, and community-leader intermediaries.

The first subtype of hierarchical digital intermediaries is political intermediaries. Although political intermediation has not been analyzed from a digital perspective, political science has extensively studied the implications of intermediation with political objectives (Valenzuela, 1977; Novaes, 2018). In many cases, this intermediation is seen as clientelistic behavior based on the concept of constituency service (Toro, 2017). Political intermediaries use their political influence to process public benefits or services in exchange for greater electoral adhesion (Hicken et al., 2016). From a digital perspective, and given the advance of e-government, digital political intermediaries can adapt their management and infrastructure to help voters access government services and benefits. In this sense, digital political intermediaries develop informal assistance networks through occasional operations or by constructing a support infrastructure in their offices. Consequently, the referral of users does not necessarily occur in spaces established to solve these concerns but can also expand through deployments mediated by electoral incentives. This political intermediation differs from bureaucratic intermediation, which involves government officials who aim to advance indicators of coverage and efficiency of public services.

The second subtype of hierarchical digital intermediaries is bureaucratic intermediaries. In a digital scenario, intermediation can be presented through the digital capacity of its bureaucracy. Public servants use their knowledge to advise which procedures can be carried out online and which digital portals should be accessed. Additionally, bureaucratic intermediaries may be responsible for mediating face-to-face procedures or receiving information when the allocation of state benefits requires background information from municipalities. Bureaucratic intermediaries can guide users in interacting with digital portals and play a complementary role in the application for and assignment of benefits when a process requires both online and face-to-face actions. This type of hierarchical intermediary is a first filter when users seek assistance with using portals and completing procedures.

The third subtype of hierarchical digital intermediaries is community-leader intermediaries. These are non-professional actors who hold leadership positions in grassroots community organizations, such as neighborhood councils or functional organizations (Liou and Stroh, 1998; Morgan-Trimmer, 2014; Shea, 2011). Their assistance has a neighbor-helping purpose beyond their official functions. In this case, knowledge about completing online procedures is self-learned. Generally, agents from neighborhood organizations possess higher levels of expertise compared to the rest of the community, and their assistance fosters perceptions of community-government reciprocity (Anglin and Herts, 2004; Daniere et al., 2005; Ionescu, Trikic and Rudas, 2024). This dynamic occurs thanks to the digital skills of these community actors who serve as a bridge and allow the government to reach users. Thus, intermediation can adopt non-clientelistic forms at the local level when the knowledge provided reflects digital reciprocity (Correa and Pavez, 2016).

**Close Digital Intermediation**

Compared to hierarchical digital intermediaries, close digital intermediaries are distinguished by the close proximity of intermediaries with users. These linkages generally are family relationships or interpersonal bonds of trust (Correa, 2014). This type of closeness often arises from sharing common spaces in everyday life. The facilitating factor of this type of intermediation lies in the user's proximity to individuals with better digital access or higher levels of digital literacy. Further, the guidance or assistance is based on interpersonal ties. Close digital intermediation is less planned than hierarchical intermediation, as it generally does not involve mobilization of resources for deployment. It is a sporadic, on-demand intermediation where the generational factor is paramount. We distinguish two subtypes of close intermediaries: family intermediaries and peer intermediaries.

Family intermediaries refer to actors who help beneficiaries due to their familial relationship with them. This group includes the beneficiaries' relatives with a higher level of knowledge or digital literacy to assist with digital procedures. Family intermediation is crucial for closing the digital gap. Younger family members have greater digital literacy to handle the challenges of technological advances (Rosen, 2010; Zur and Zur, 2011), and their digital literacy has become pivotal in addressing the age-related digital gap. Children may help their families use technology for different purposes, including solving problems and integrating into the digital world (Correa, 2024). Undoubtedly, family intermediation is crucial for enhancing digital interactions in health, education, and employment (Choudrie, Ghinea, and Songonuga, 2013; Jun, 2020; Darmody et al. 2022; Lam and Lee, 2005; Walker, 2017).

On the other hand, peer intermediaries are individuals who, like family intermediaries, have a personal relationship with users, but instead of being relatives, peer intermediaries are often neighbors and acquaintances with better digital access or greater digital literacy. Given the existing connectivity and digital literacy gaps, these intermediaries, in lending their skills and internet access to their less advantaged acquaintances, serve as essential bridges.

**Digital Intermediation in Chile: Testing the Typology in a Crucial Case**

Chile, a 19-million people South American country, is a crucial case for analyzing the typology of this article due to the high levels of internet penetration and advanced e-government amidst strong socioeconomic and digital inequalities. Household Internet access increased from 70.2% in 2015 to 94.3% in 2023 (Subtel, 2023). According to official government data for 2023, 89% of government procedures are digital.[3] In this context, digitalization not only solves various institutional and social problems through technology, but it also obscures its own political and contestable nature (Collington, 2022). Interactions between citizens and digital states like Chile provide opportunities to explore the hidden linkages precipitated by e-government (Datta, 2023 ).Digital growth in Chile reached a significant milestone during the COVID-19 pandemic when the country, through the Digital Government unit, succeeded in massifying a unique digital identity called the *ClaveÚnica* (UniqueKey). This system creates a personal and non-transferable digital identity, linking the identification number to a key that allows access to various government platforms and services. This unique access has precipitated a massive migration of users from physical to digital interactions. Currently, Chile enables digital access to multiple government services and the completion of online procedures with security, speed, and efficiency (Universidad

de Chile, 2023). The ClaveÚnica in Chile allows individuals to remotely apply for government benefits and subsidies, as well as complete other bureaucratic procedures such as obtaining certificates and filing complaints (Government of Chile, 2017). This mechanism is efficient for the Chilean government, as it reduces transaction costs by minimizing inherently more costly face-to-face services in physical offices.

The ClaveÚnica digital identity is one of the main tools in Chile for improving efficiency and modernizing public procedures. However, despite Chile's advanced digitalization, Chile faces the predicament of having almost full internet penetration while the population experiences difficulties using these platforms. The high internet penetration and government digitalization can obscure significant disparities in internet quality, skills and usage. Furthermore, geographic and socioeconomic factors leave some communities without reliable, high-speed connectivity. A recent report on digitalization at the district level in Chile highlights persistent gaps in connectivity, digital resources in schools, and access to e-government services across various regions (NUDOS, 2024). At the same time, Chile has some of the poorest general literacy and digital literacy skills among OECD countries (OECD, 2021). It is precisely this poor digital literacy among Chileans that has generated gaps and biases that have not yet been fully addressed (Correa. Valenzuela and Pavez, 2024).

*1) A descriptive analysis of the ClaveÚnica in Chile.*

The obstacles to popular digital literacy combined with the government's intense digital deployment and simultaneous decrease in face-to-face services make an excellent empirical scenario to observe digital intermediaries. To analyze digital intermediation in Chile, we conducted the 2023 Digital Inclusion Survey. This face-to-face survey was administered between August and September 2023 to permanent residents over 14 years old in communes selected by population size using differentiated sampling mechanisms for rural and urban areas. For urban areas, clusters were established by block, while for rural areas, clusters were established using a simple random route.

A probabilistic sampling design was carried out by commune, although not proportional at the regional level, with a sampling error of ±2.8% and simple random sampling, maximum variance, and 95% confidence. Interviews were conducted in both urban and rural settings, with one individual randomly selected from each household. The final sample includes 1,200 respondents, comprising 751 from urban areas and 449 from rural communities. The sample comprises more women (57.4%) than men (42.6%), with a median age of 52 years. and the median educational level is secondary education.

Part of the questionnaire was designed to measure knowledge of the government's platforms, the intensity of use of digitalized services, and the existence of support networks for access. The questions allow an important cross-checking of variables and analysis of the conditions that influence the use of e-government.

We believe that identifying these intermediaries in the digital distribution of public services helps to qualify the notion that digital procedures are more direct and avoid involvement of third parties. For example, when asked about the frequency of use of the ClaveÚnica to complete digital procedures with the government, a significant percentage of the sample indicated that they had never used the ClaveÚnica for such procedures, despite its widespread coverage and the fact that it is a requirement for these procedures.

A descriptive analysis of Clave Única in Chile

Table 1
Use of the ClaveÚnica for digital procedures with the state

|  | Freq. | Percent |
|---|---|---|
| Never | 548 | 45.67 |
| Not used for over three months | 199 | 16.58 |
| Between one and three times a month | 308 | 25.67 |
| Every week | 108 | 9.00 |
| Every day | 27 | 2.25 |
| Several times a day | 1 | 0.08 |
| Not sure/no response | 9 | 0.75 |
| Total | 1,200 | 100.00 |

The results are surprising because official Chilean government data shows that the coverage of ClaveÚnica is almost complete. The paradox is that, despite total coverage, almost 46% of respondents stated that they have never used it for e-government procedures, while nearly 54% stated that they use it, indicating a significant entry barrier to e-government usage. Other results are slightly more promising: 25.6% stated that they use the ClaveÚnica one to three times a month, followed by 16.5% who stated that they have not used it for over three months. More frequent users total 11.3%, with 9% using the ClaveÚnica weekly, 2.2% using it daily, and 0.08% using it several times per day (0.08%).

Beyond uneven use, the usage frequency of ClaveÚnica is also unevenly concentrated among those who use it sporadically, possibly for specific procedures. This heterogeneity could be a consequence of poor digital literacy among non-users, as shown in table 1.

Table 2
Does anyone in your household know someone who can help you with the use of technology (computer, internet, etc.)?

|  | Freq. | % |
|---|---|---|
| Yes, only someone outside the home | 380 | 31.67 |
| Yes, only someone inside the home | 321 | 26.75 |
| Both inside and outside the home | 312 | 26.00 |
| No, do not know anyone | 176 | 14.67 |
| Not sure/no response | 11 | 0.92 |
| Total | 1,200 | 100.00 |

Source: Prepared by authors based on data from NUDOS 2023 Digital Inclusion Survey.

The survey classifies those who seek help to complete digital procedures into three categories: have help only from people inside the home, have help only from people outside the home, or have help from people both inside and outside the home. Approximately 32% have help only from people outside the home, while 27% have help only from people inside the home, and 26% have help from people both inside and outside the home. These results demonstrate a reliance on and trust in others to complete digital procedures. Moreover, even those who lack help from people inside the home find assistance from people outside the home, highlighting the importance of support networks and illustrating how close intermediation operates.

*2) Testing the Typology: A Factor Analysis.*

To test the typology of intermediaries, the Digital Inclusion Survey included a matrix question about people who are likely to act as digital intermediaries for others. Using a dichotomous scale ("yes" or "no"), it asked from whom the respondent had sought help to complete digital procedures with the government among a list of 14 types of intermediaries, "others" (not listed), and "no one." The question was presented as follows:

| E3. Have you asked any of the following people for help to complete digital government procedures? | | |
|---|---|---|
| *(Each item below listed individually) | (1) Yes | (2) No |
| *E3_1 Parent; E3_2 Spouse/partner; E3_3 Son/daughter; E3_4 Other relative; E3_5 Friend; E3_6 Work colleague; E3_7 Teacher; E3_8 Neighbor; E3_9 Neighborhood council member or community leader; E3_10 Public official (examples: social worker, municipal secretary, among others); E3_11 Municipal councilor (or person working for him/her); E3_12 Mayor (or person working for him/her); E3_13 Regional governor (or person working for him/her); E3_14 Deputy or senator (or person working for him/her); E3_15 Other; E3_16 No one | | |

In terms of descriptive results, there is a predominance of people who state that they ask for help from people close to them or family members, particularly their child or other relatives, friends, and neighbors. In this sense, digital applications for public benefits are highly mediated by close links, which is a type of close intermediation. Interestingly, although in smaller numbers, a significant group also declares hierarchical digital intermediation by public officials or political representatives. The following table shows the distribution of responses by type of intermediary (see Table 3).

Table 3 Who have you asked for help to complete digital government procedures?

| Intermediaries | Yes |
| --- | --- |
| E3_3 Son/daughter | 42% |
| E3_4 Other relative | 25% |
| E3_5 Friend | 23% |
| E3_8 Neighbor | 20% |
| E3_2 Spouse/partner | 20% |
| E3_6 Work colleague | 13% |
| E3_10 Civil servant | 7% |
| E3_1 Parent | 7% |
| E3_7 Teacher | 6% |
| E3_9 Neighborhood council member or community leader | 4% |
| E3_11 Municipal councilor | 2% |
| E3_13 Regional governor | 2% |
| E3_12 Mayor | 1% |
| E3_14 Deputy or senator (or person working for him/her) | 1% |
| E3_15 Other | 2% |

Source: Prepared by authors based on data from the 2023 Digital Inclusion Survey.

Next, to test the presence of our theoretical dimensions, we performed an confirmatory factor analysis on the fourteen types of agents identified in the survey. We believe that factor analysis is an optimal technique to corroborate our theoretical dimensions on observational data because variables can be congregated into common patterns or underlying dimensions that are not directly observable Thus, it allows us to reduce variables using grouping.

These dimensions consist of variables that are homogeneous among themselves and heterogeneous from one another. Although there are several types of rotation, in this article we use orthogonal rotation (Varimax), since it allows grouping with minimal shared variance among the factors and thus generates more independence of the factors. This reduces the number of variables with high saturations in each factor (Lloret-Segura et al., 2014). The results of factor analysis rotated by maximum variance are as follows:

Table x: title

| Variable | Factor1 | Factor2 | Factor3 | Factor4 | Factor5 | Factor6 | Uniqueness |
|---|---|---|---|---|---|---|---|
| parent | 0.2295 | 0.4407 | 0.0617 | 0.0245 | -0.2623 | -0.0468 | 0.6777 |
| spouse/partner | 0.1522 | 0.3411 | 0.0085 | 0.3337 | 0.0621 | -0.0644 | 0.7411 |
| child | 0.0488 | 0.0379 | 0.1160 | 0.0371 | 0.3084 | -0.0166 | 0.8860 |
| other relative | 0.1361 | 0.3679 | 0.0799 | -0.0993 | 0.1261 | -0.0616 | 0.8102 |
| friend | 0.1081 | 0.6494 | 0.0259 | 0.0817 | -0.0094 | 0.0109 | 0.5591 |
| colleague | 0.1740 | 0.3898 | -0.0074 | 0.2902 | -0.0502 | 0.0764 | 0.7251 |
| teacher | 0.3603 | 0.2814 | 0.1465 | -0.0820 | -0.1843 | 0.1457 | 0.7076 |
| neighbor | 0.1184 | 0.5439 | 0.0850 | -0.0422 | 0.0482 | 0.0162 | 0.6785 |
| neighborhood leader | 0.5179 | 0.1324 | 0.3187 | 0.0407 | 0.0228 | -0.0532 | 0.6076 |
| officials | 0.3793 | 0.1608 | 0.3230 | -0.0338 | 0.0318 | 0.0680 | 0.7192 |
| councilors | 0.6156 | 0.1305 | 0.0410 | -0.0773 | 0.0487 | 0.1209 | 0.5793 |
| mayors | 0.7432 | 0.0596 | -0.0090 | 0.0250 | 0.0286 | 0.0418 | 0.4409 |
| regional authorities | 0.6196 | 0.0678 | -0.0510 | 0.0057 | -0.0708 | -0.0990 | 0.5940 |
| congressmen | 0.6792 | 0.1514 | 0.1295 | 0.1082 | -0.0681 | -0.0450 | 0.4806 |

Using the results of the factor analysis, figure 2 below visualizes the differences in the variances of the two main types of digital intermediaries, which are hierarchical and close. The vertical axis identifies hierarchical intermediation, and the horizontal axis identifies close intermediation. Furthermore, within each of these two main dimensions, variances may reveal relationships between subtypes. For example, in hierarchical intermediation, there is a high correlation between political intermediaries such as parliamentarians and local authorities. It is also interesting to observe bureaucratic intermediaries as a group differentiated from political and community intermediaries. Meanwhile, close intermediation identifies the family and peer subtypes. While partners, immediate family, and parents are grouped in a subdimension, friends and neighbors are statistically distanced. Notably, the son/daughter option does not correlate with either of the two main dimensions, which suggests the existence of a generational factor that may be cross-sectional. This is consistent with the literature discussed above showing that early technological exposure allows younger individuals to help family members and neighbors with technological difficulties.

Figure 2. Factor analysis visualization graphic

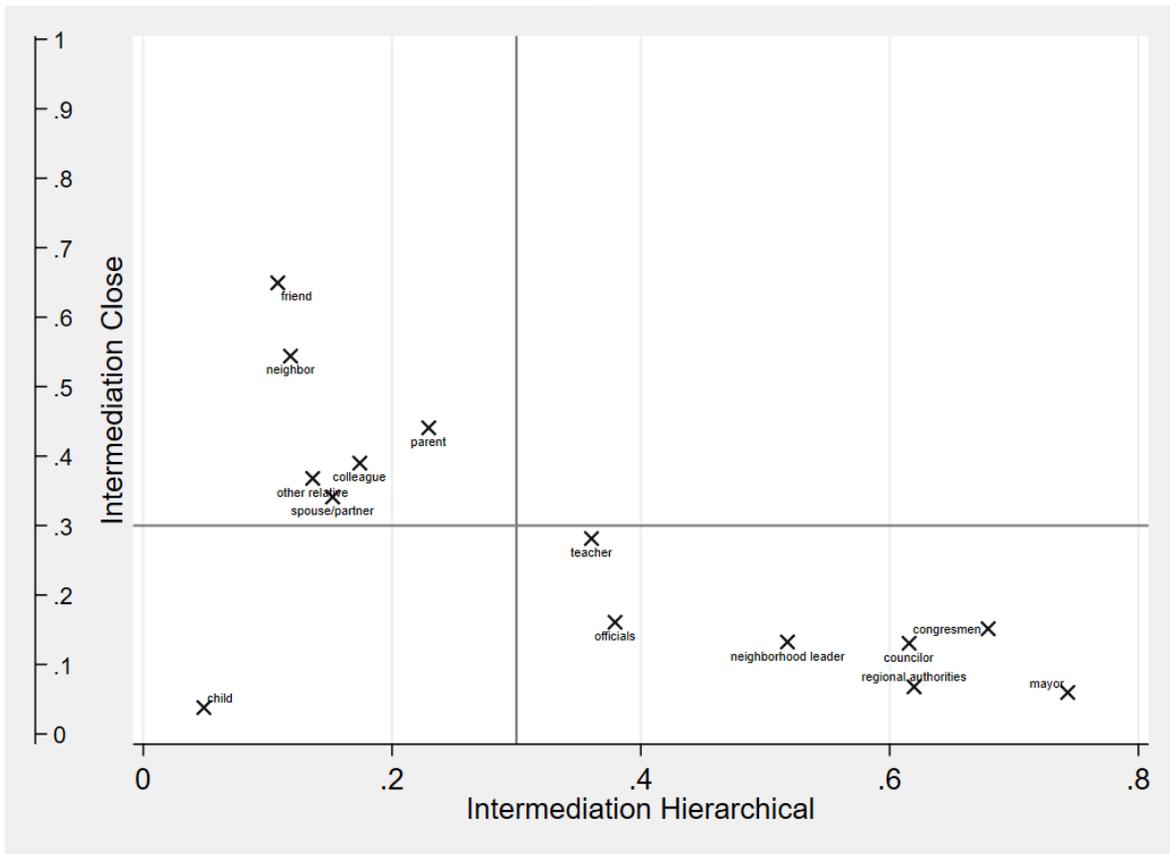

*Source*: Prepared by authors based on data from NUDOS 2023 Digital Inclusion Survey.

**Concluding Remarks**

Government digitalization is a complex process that has transformed the physical and logistical aspects of bureaucratic procedures for citizens as well as internal government processes. Government digitalization is widely lauded for the advantages it brings to individuals and governments (Lenk and Traunmuller, 2000). However, the challenges on society as a whole are often overlooked. More specifically, government digitalization has profoundly changed the structure and dynamics of the citizen-government relationship. The traditional model saw citizens interacting with the government through street-level bureaucrats, but that paradigm is being eradicated. Through e-government, bureaucrats have been replaced by digital procedures and automated systems, leading to a depersonalized relationship between citizens and their governments.

There is also some concern that, due to the connectivity and digital literacy gaps among citizens, the advantages of government digitalization are inequitably distributed. That is, digitalization separates people with the economic resources to access digital devices and internet connectivity from those who do not. Likewise, digitalization also separates people who have the education and skills to effectively navigate e-government from those who do not. The result is that more-advantaged individuals receive more advantages from digitalization, while less-advantaged individuals receive less advantages from digitalization—a result that is neither equitable nor desirable.

The current approach to government digitalization has created an unintended paradox: while more advanced technological processes and infrastructures are being developed, they are often inaccessible to individuals lacking both the necessary digital access and skills to effectively use these platforms. As a result, the very digital tools intended to enhance government reach and service inclusivity may inadvertently exclude the citizens they aim to serve. Understanding digital intermediaries is important because, among other reasons, societies are experimenting a challenging restructuring and transformation of relationships and social interactions. While individuals and groups are connecting through social networks and digital platforms, governments are still transforming their processes of digitalization. Both society and the government must sort out these challenges and existing digital gaps in order to realize the stated intentions of government digitalization. To this end, these real-world operational gaps that have emerged present an opportunity for scholars and practitioners to understand what is missing conceptually. In particular, these gaps allow us to discover digital intermediaries as a new political concept to fill this conceptual void.

Although there is previous research indicating that lower digital skills lead to digital gaps (Correa et al. 2024) with intermediaries emerging as agents of technological socialization. These intermediaries help bridge the gap between end users and digital services by providing assistance to those withough digital access or literacy (DESUC 2017). Thus, effectively classifying digital intermediaries is a critical issue for both scholars and practitioners seeking to understand digital inequalities and public services. A typology of these actors enables us to understand under what contexts they operate, and whether their action can be understood from within or outside institutional initiatives. As demonstrated in this article, there are various types of digital intermediaries that can be grouped into hierarchical and close types. Our typology suggests observing these actors according to their relationship with beneficiaries and their institutional or extra-institutional position.

This article aims to propose a theoretical typology of digital intermediaries to enhance the discussion on the implications of government digitalization on the relationships between citizens and their governments. For this reason, the scope of this article is theoretical as it classifies the actors involved in the digital process. Although it is an empirical verification of the proposed dimensions, this article does not seek to find the reasons for this intermediation, nor to explain the factors that have influenced decisions to seek help completing digital government procedures.

Although e-government offers several benefits for the administration of public services and makes the resolution of certain needs more agile, existing social inequalities cause unaddressed digital gaps. In this void, digital intermediaries become critical agents within social environments. As a result, e-government as an instrument of public action may have unintended social consequences and lead to dependence on digital intermediaries. Moreover, these intermediaries are not always everyday or close links; they are also political or bureaucratic. Thus, each intermediary acts under different social codes and incentives. It is this complexity that calls for new research into digital intermediaries.

**Endnotes**

1. The authors acknowledge funding from Chile's National Agency of Research and Development (ANID) through Fondecyt 1210740 and the Millennium Nucleus on Digital Inequalities and Opportunities (NUDOS) [grant NCS2022_046].

2. Internet Society. "2023 Year in Review." Internet Society, 2023. Accessed January 10, 2024. https://www.internetsociety.org/wp-content/uploads/2024/03/2023-Year-In-Review-EN.pdf.

3. Secretaría de Gobierno Digital, "Transformación Digital del Estado al Servicio de las Personas," Gobierno Digital, accessed October 11, 2024, http://digital.gob.cl/.